\documentclass[fleqn]{article}

\usepackage{graphicx}

\def\dsl#1{{#1}\!\!\!/}

\begin{document}

\title{
{\sf Extended BRS Symmetry and Gauge Independence in On-Shell
Renormalization Schemes}
}
\author{ S. Alavian and T.G. Steele\thanks{email: Tom.Steele@usask.ca}\\
{\sl Department of Physics and Engineering Physics }\\
{\sl University of Saskatchewan}\\
{\sl Saskatoon, Saskatchewan S7N 5E2, Canada.}
}
\maketitle
\begin{abstract}
Extended BRS symmetry is used to prove gauge independence of the 
fermion renormalization constant $Z_2$ in on-shell QED  renormalization schemes.
A necessary condition for gauge independence of $Z_2$ in on-shell  QCD 
renormalization schemes is formulated.  Satisfying this necessary  condition appears to be 
problematic at the three-loop level in QCD.
\end{abstract}
In on-shell schemes, the fermion mass renormalization $Z_m$ and wave function 
renormalization  
$Z_2$ have been observed to 
be gauge parameter independent in explicit two-loop QED and QCD calculations 
\cite{bgs}.   
Gauge parameter independence of $Z_2$ is phenomenologically significant 
because it implies that the difference between the (fermion) anomalous dimension of 
heavy quark effective theories 
\cite{iw} and  QCD is gauge independent.  

An extension of BRS symmetry, which allows variations of the gauge parameter 
to be included as part of the symmetry transformations \cite{ps}, will be applied  to  the
gauge parameter dependence of $Z_2$.  
This approach results in an extension of Slavnov-Taylor identities,  allowing 
gauge dependence to be formulated algebraically. 
Previous application of these techniques resulted in a proof of
the gauge independence of the mass renormalization $Z_m$ to all orders
in on-shell QED and QCD renormalization schemes \cite{bls}.  We will prove the gauge parameter 
independence of $Z_2$ in 
on-shell schemes for QED
and formulate a necessary condition for gauge independence in QCD which appears 
problematic beyond the two-loop level.
This  complements earlier work on 
gauge independence of $Z_2$ in QED resulting in the (dimensionally-regularized)
relation \cite{z2}
\begin{equation}
\frac{\partial Z_2}{\partial \xi}\sim \int d^Dk\frac{1}{k^4}=0
\label{jz}
\end{equation}
where $\xi$ is the gauge parameter and the massless tadpole is zero in 
dimensional regularization.   Since the QED result (\ref{jz}) cannot be extended to QCD, our 
extended BRS symmetry proof for QED provides a new approach to formulating 
questions of gauge independence  of $Z_2$ in QCD.

The QED Lagrangian  in the auxiliary field formalism \cite{nl} for covariant gauges is
\begin{equation}
{\cal L}=-\frac{1}{4}F^2+\bar \psi\left(i\dsl{D}-m\right)\psi +\frac{\xi}{2} B^2 +B\partial 
\cdot A-\bar c\partial^2 c
\label{l_qed}
\end{equation}
where $F$ is the field strength and $B$ is the auxiliary gauge
field.
This Lagrangian is invariant under the  BRS symmetry 
\begin{eqnarray}
& & \delta A_\mu=\epsilon \partial_\mu c\quad ,\quad \delta \bar\psi=i\epsilon g c\bar 
\psi\quad ,\quad \delta c=0\nonumber\\
& &\delta B=0\quad ,\quad \delta \psi=-i\epsilon g c\psi \quad ,\quad \delta \bar c=0
\label{qed_brs}
\end{eqnarray}
where $\epsilon$ is a global grassmann quantity.
The auxiliary field formalism guarantees nilpotence of the BRS transformations 
without invoking equations of motion.

An extension of BRS symmetry that includes gauge parameter variations introduces a
new   term in the Lagrangian  
\begin{equation}
{\cal L}\rightarrow {\cal L}+\frac{\chi}{2}\bar c B
\label{ex_l_qed}
\end{equation}
where $\chi$ is a global grassmann variable.  
Although $\chi$ will be set to zero after functional differentiation, it is still important 
to recognize that since $\chi$ is a global 
Grassmann quantity, it does not change the dynamics of any process with zero ghost number.
The modified Lagrangian (\ref{ex_l_qed}) is invariant under the following extended 
BRS symmetry \cite{ps}
\begin{eqnarray}
& &\delta^+ A_\mu=\epsilon \partial_\mu c\quad ,\quad \delta^+ \bar\psi=i\epsilon g c\bar 
\psi\quad ,\quad \delta^+ c=0\nonumber\\
& &\delta^+ B=0\quad ,\quad \delta^+ \psi=-i\epsilon g c\psi \quad ,\quad \delta^+ \bar c=B
\label{qed_x_brs}\\
& &\delta^+\xi=\epsilon\chi\quad ,\quad \delta^+\chi=0
\end{eqnarray}

As for BRS symmetry, the extended BRS symmetry (\ref{qed_x_brs}) implies the the 
following relation for the effective action $\Gamma$.
\begin{equation}
0=\partial_\mu c \frac{\delta \Gamma}{\delta A_\mu}
+ \frac{\delta\Gamma}{\delta \bar K} \frac{\delta \Gamma}{\delta \psi}
+  \frac{\delta\Gamma}{\delta K}
\frac{\delta \Gamma}{\delta \bar\psi}
+  B\frac{\delta \Gamma}{\delta \bar c}
+\chi\frac{\partial\Gamma}{\partial\xi}
\label{qed_gamma_x_brs}
\end{equation}
where $K$ is a current coupled to the composite operator $\delta^+\bar \psi$ and 
$\bar K$ is coupled to $\delta^+ \psi$.
Differentiating (\ref{qed_gamma_x_brs}) with respect to $\chi$, $\bar\psi(x)$,
$\psi(y)$, setting $\chi=0$ and imposing ghost number conservation leads to 
the following identity for the proper fermion two-point function \cite{bls}.
\begin{equation}
\frac{\partial}{\partial\xi}\frac{\delta^2\Gamma}{\delta\psi(y)\delta
\bar\psi(x)}  =
+
\frac{\delta^3\Gamma}{\delta\psi(y)\delta\bar K\delta \chi}
\frac{\delta^2\Gamma}{\delta\bar\psi(x)\delta\psi}
+
\frac{\delta^2\Gamma}{\delta\psi(y)\delta\bar\psi}
\frac{\delta^3\Gamma}{\delta\bar\psi(x)\delta K\delta\chi}
\label{gamma_ident}
\end{equation}
Transforming to momentum space and defining
\footnote{An implicit coordinate integration is associated with the $\chi$ derivative.}
\begin{eqnarray}
& &\frac{\delta^2\Gamma}{\delta\chi\delta {\bar K}(w)\delta\psi (y)}
=\int \frac{d^4q}{(2\pi)^4}\frac{d^4\ell}{(2\pi)^4}\,e^{-iq\cdot(y-z)
-i\ell\cdot(w-z)}
F(q,\ell,-q-\ell)\label{F}\\
& &
\frac{\delta^2\Gamma}{\delta\chi\delta K(w)\delta{\bar\psi} (y)}
=\int \frac{d^4q}{(2\pi)^4}\frac{d^4\ell}{(2\pi)^4}\,e^{-iq\cdot(x-z)
-i\ell\cdot(w-z)}
{\bar F}(q,\ell,-q-\ell) \label{Fbar}
\end{eqnarray}
results in the final form needed for studying the gauge dependence of the 
fermion propagator $S_F$ in QED \cite{bls}.
\begin{equation}
\frac{\partial}{\partial\xi}S_F^{-1}(p)=S_F^{-1}(p)\left[ F(p,-p,0)+
\bar F(-p,p,0) \right]
\label{prop_ident}
\end{equation}
Note that the Green functions $F(p, -p, 0)$  and $\bar F(p, -p, 0)$ 
cannot have single particle poles.

In on-shell renormalization schemes  the bare mass $m_0$ and 
the renormalized mass $M$ are related through the condition
\begin{equation}
\biggl. S_F^{-1}(p)\biggr|_{\dsl{p}=M}=0
\label{mass_shell}
\end{equation}
This results in the definition of the mass renormalization constant.
\begin{equation}
\frac{m_0}{M}=Z_m
\label{Z_m}
\end{equation}
The wave function renormalization constant $Z_2$ is the residue 
of $S_F$ at the $\dsl{p}=M$ pole.
\begin{equation}
Z_2=\lim_{\dsl{p}=M} \left(\dsl{p} -M\right)S_F(p)
\label{Z_2}
\end{equation}
Perturbative expansions of $Z_m$ and $Z_2$ have been calculated to two-loop order
in a scheme which dimensionally regulates both the infrared and ultraviolet 
divergences, resulting in explicitly gauge independent expressions 
for QED and QCD \cite{bgs}.  

The mass renormalization $Z_m$, and hence $M$, has been proven to be gauge 
independent to all orders of perturbation theory \cite{bls,kron}. 
Thus when  both sides of (\ref{prop_ident}) are divided by $\dsl{p}-M$ 
the quantity $\dsl{p}-M$ commutes with the $\xi$ derivative.
\begin{equation}
\frac{\partial}{\partial\xi}\left(\frac{S_F^{-1}(p)}{\dsl{p}-M}\right)
=\frac{S_F^{-1}(p)}{\dsl{p}-M}
\left[ F(p,-p,0)+\bar F(-p,p,0) \right]
\label{Z_2F1}
\end{equation}
Using the property that
\begin{equation}
S_F^{-1}(p)=\frac{\dsl{p}-M}{Z_2}+{\cal O}
\left[\,\left(\dsl{p}-M\right)^2\,\right]
\label{S_F_property}
\end{equation}
along with the gauge independence of $M$ leads to the following 
result when (\ref{Z_2F1}) is evaluated on-shell.
\begin{equation}
\frac{\partial}{\partial\xi}\left(\frac{1}{Z_2}\right)=\frac{1}{Z_2} 
\lim_{\dsl{p}=M}
\left[
F(p,-p,0)+\bar F(-p,p,0)\right]
\label{Z_2_rsult}
\end{equation}
This is our central result for QED:  the gauge dependence of the wave function renormalization constant is related 
to the on-shell properties of the Green function
$F(p,-p,0)+\bar F(p,-p,0)$.  In particular, if this Green function is zero on-shell, then $Z_2$ is gauge independent.

Before studying the on-shell behaviour of $F(p,-p,0)+\bar F(p,-p,0)$ we review some aspects of the auxiliary field formalism.    
Since the $B$ field and $\partial\cdot A$ are mixed in the Lagrangian 
(\ref{l_qed}) the quadratic part of the Lagrangian must be diagonalized, leading to the
free field propagators
\begin{eqnarray}
& &
\int d^4x\,e^{ip\cdot x}\langle O\vert T\left(B(x) B(0)\right)\vert O\rangle
=0 \label{bb_prop}\\
& &
\int d^4x\,e^{ip\cdot x}\langle O\vert T\left(B(x) A_\mu(0)\right)\vert O
\rangle = \frac{p_\mu}{p^2}\equiv G_\mu(p) \label{ba_prop}\\
& &
\int d^4x\,e^{ip\cdot x}\langle O\vert T\left(A_\mu(x) A_\nu(0)\right)
\vert O
\rangle =
i\left[-\frac{g^{\mu\nu}}{p^2} +(1-\xi)\frac{p^\mu p^\nu}{p^4}\right] 
\equiv D^{\mu\nu}(p)
\label{aa_prop}
\end{eqnarray}
BRS symmetry implies that (\ref{bb_prop}) and (\ref{ba_prop}) are valid to all orders 
in perturbation theory \cite{bls}.

\begin{figure}
\centering
\includegraphics[scale=0.7]{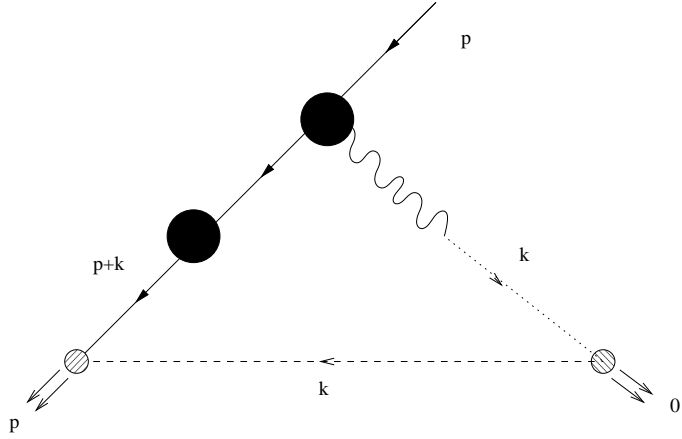}
\caption{Feynman diagram expressing $F(p,-p,0)$ in terms of  one-particle irreducible functions 
represented by the solid circles. Dashed lines represent the ghost field, 
and the dotted line represents the auxiliary field $B$.  Composite operators 
coupled to the currents are represented by the
partially-filled circles.}
\label{f_fig}
\end{figure}

As illustrated in Figure \ref{f_fig}, the (QED) Green function $F(p,-p,0)$ is easily written in terms of
one-particle irreducible Green functions 
\begin{equation}
F(p, -p, 0)=\int d^Dk \,\Gamma_\mu(k, p) G_\mu(k) S_F(p+k) \tilde D(k^2)
\label{on-shell_F_1}
\end{equation}
where $\tilde D(k^2)$ is the ghost propagator (which for QED corresponds to the free  
field result) and the fermion-photon vertex function $\Gamma_\mu$ is defined by
\begin{equation}
S_F(p)\Gamma_\nu (p, k) S_F(p+k) D^{\mu \nu}(k)
=\int d^Dx \int d^Dy \,\,e^{i k\cdot x+i p\cdot y}\langle O\vert T\left[
\psi(0) A_\mu(x) \bar \psi (y)
\right]\vert O\rangle
\label{vertex}
\end{equation}
Substituting (\ref{ba_prop}) and the (free-field) ghost propagator into (\ref{on-shell_F_1}) and
using the Ward identity for the vertex function 
\begin{equation}
k^\mu \Gamma_\mu(p,k)=S_F^{-1}(p+k)-S_F^{-1}(p)
\label{ward}   
\end{equation}
simplifies the expression for $F(p, -p, 0)$.
\begin{equation}
F(p,-p,0)=iS_F^{-1}(p) \int d^Dk \frac{1}{k^4} S_F(p+k)-i\int d^Dk\frac{1}{k^4}
\label{on-shell_F_2}
\end{equation}
The second term in the above equation is  a massless tadpole which is zero in dimensional
regularization, leading to the final expression for $F(p, -p, 0)$ in QED.
\begin{equation}
F(p,-p,0)=iS_F^{-1}(p) \int d^Dk \frac{1}{k^4} S_F(p+k)
\label{on-shell_F_3}
\end{equation}
In the on-shell scheme \cite{bgs} 
infrared and ultraviolet divergences are dimensionally regulated, so 
the integral in (\ref{on-shell_F_3}) is finite on-shell. Thus the 
$S_F^{-1}(p)$ prefactor in
(\ref{on-shell_F_3}) implies that $F(p, -p, 0)$ is zero at the $\dsl{p}=M$ 
mass-shell. This argument can be trivially extended to $\bar F(p, -p, 0)$, 
and we conclude that to all orders in QED
\begin{equation}
\biggl. F(p, -p, 0)+\bar F(p, -p, 0) \biggr|_{\dsl{p}=M}=0
\end{equation}
and hence from the result (\ref{Z_2_rsult}) we have proven the gauge independence
of the QED renormalization constant $Z_2$ in mass-shell schemes.

\begin{figure}
\centering
\includegraphics[scale=0.7]{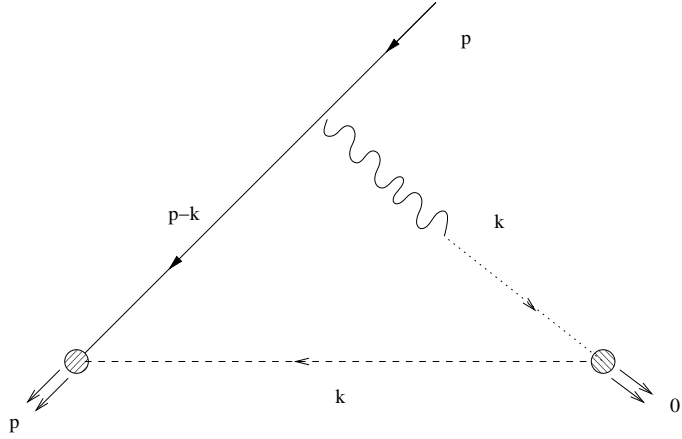}
\caption{Feynman diagram for one-loop contributions to  $F(p,-p,0)$. 
Dashed lines represent the ghost field, 
and the dotted line represents the auxiliary field $B$.  Composite operators 
coupled to the currents are represented by the
partially-filled circles.}
\label{f_1l_fig}
\end{figure}

An explicit illustration of the on-shell behaviour of $F(p, -p,0)+\bar F(p, -p, 0)$
in the regularization scheme \cite{bgs} to one-loop order requires evaluation of the  
diagram in Figure \ref{f_1l_fig}.  In terms of the integrals
(with the convention $D=4+2\epsilon$)
\begin{eqnarray}
& &\int \frac{d^Dk}{(2\pi)^D} 
\frac{1}{\left[(k-p)^2-m_0^2\right]^\alpha \,k^{2\beta}}
=I\left[\alpha, \beta\right]
\label{I_alpha_beta}\\
& &\int \frac{d^Dk}{(2\pi)^D} 
\frac{k^\mu}{\left[(k-p)^2-m_0^2\right]^\alpha \,k^{2\beta}}=p^\mu 
J\left[\alpha , \beta\right]
\label{J_alpha_beta}
\end{eqnarray}
we find the one-loop expression for $F(p,-p,0)+\bar F(p,-p,0)$.
\begin{equation}
F(p, -p, 0)+\bar F(p, -p, 0)=2i g^2\left[ m_0 \left(\dsl{p}-m_0\right)
J(1,2)+\left(p^2+m_0^2\right) J(1,2) -I(1,1)\right]
\label{F+bar_F}
\end{equation}
and hence the on-shell behavior of $F+\bar F$ to one-loop order is given by
\begin{equation}
\lim_{\dsl{p}=M} \left[F(p,-p,0)+\bar F(p,-p,0)\right]=2ig^2
\lim_{\dsl{p}=M=m_0} \left[ 
2m_0^2 J(1,2)-I(1,1)\right]
\label{on_shell_F}
\end{equation}
The desired on-shell values for the integals in (\ref{on_shell_F}) can be 
reduced to evaluation of a single class of scalar integrals.
\begin{equation}
\Lambda\left[\alpha, \beta\right] =
\int \frac{d^Dk}{(2\pi)^D} 
\frac{1}{\left[k^2+2 p\cdot k\right]^\alpha \,k^{2\beta}}
\end{equation}
a particular example being a relation between $J(\alpha, \beta)$ and
$\Lambda(\alpha, \beta)$
\begin{equation}
\lim_{\dsl{p}=m_0}J(\alpha, \beta)=\frac{1}{2 m_0^2}\left[
\Lambda(\alpha, \beta-1) -\Lambda(\alpha-1, \beta)\right]
\end{equation}
The integration by parts technique \cite{ct} for these on-shell integrals leads
to recursion relations among the $\Lambda(\alpha, \beta)$.  The identities
\begin{eqnarray}
& &0=\int d^Dk \frac{\partial}{\partial k^\mu}\left(
\frac{p^\mu}{\left[k^2+2 p\cdot k\right]^\alpha \,k^{2\beta}}
\right)\label{int_by_parts_1}\\
& &0=\int d^Dk \frac{\partial}{\partial k^\mu}\left(
\frac{k^\mu}{\left[k^2+2 p\cdot k\right]^\alpha \,k^{2\beta}}
\right)\label{int_by_parts_2}
\end{eqnarray}
lead to the recursion relations
\begin{eqnarray}
& &0= -\beta \Lambda(\alpha-1, \beta+1)+(\beta-\alpha)\Lambda(\alpha, \beta)-2\alpha m_0^2
\Lambda(\alpha+1, \beta)+\alpha \Lambda(\alpha+1, \beta-1)
\label{rec_1}\\
& & 0=(D-2\beta-\alpha)\Lambda(\alpha, \beta)-\alpha\Lambda(\alpha+1, \beta-1 )
\label{rec_2}
\end{eqnarray}
The recursion relation (\ref{rec_2}) can also be obtained from 
dimensional analysis.
These recursion relations allow the on-shell behaviour of the one-loop integrals, 
after setting mass tadpoles to zero, to be reduced to
the fundamental dimensional regularization result
\begin{equation}
\Lambda(\alpha, 0)=\int\frac{d^D k}{(2\pi)^D} \frac{1}{\left[ k^2-m_0^2\right]^\alpha}
=\frac{i}{(4\pi)^{D/2}}\left(-m_0^2\right)^{2-\alpha} m_0^{2\epsilon}
\frac{\Gamma(\alpha-2-\epsilon)}{\Gamma(\alpha)}
\label{fund_dim_reg}
\end{equation}
Using the above techniques it is simple to find the on-shell integrals required in
(\ref{on_shell_F}).
\begin{eqnarray}
& & J(1,2)=\frac{i}{(4\pi)^{D/2}}m_0^{2\epsilon} \frac{\Gamma(-\epsilon)}{2 m_0^2(D-3)}
\label{J(1,2)}\\
& & I(1,1)=\frac{i}{(4\pi)^{D/2}}m_0^{2\epsilon} \frac{\Gamma(-\epsilon)}{(D-3)}
\end{eqnarray}
and hence in the on-shell regularization scheme \cite{bgs}, the Green function
$F+\bar F$ is zero on-shell to one-loop order,  providing a specific example of our general
result.

\begin{figure}
\centering
\includegraphics[scale=0.7]{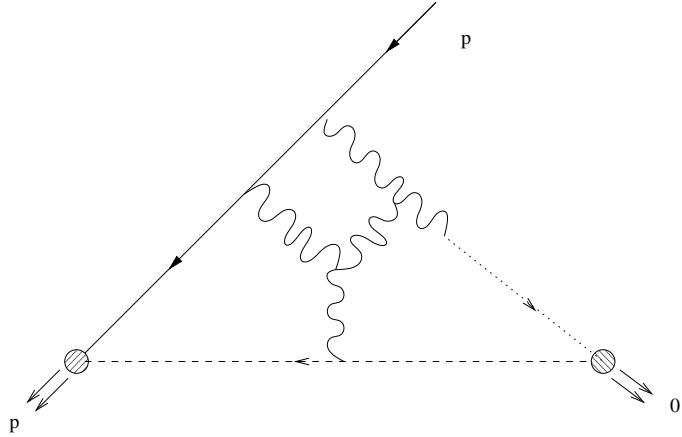}
\caption{A three-loop QCD diagram contributing to $F(p,-p,0)$ which cannot be reduced to 
the the form (\protect\ref{on-shell_F_1}) composed of fundamental one-particle irreducible 
Green functions. 
}
\label{non_ab_fig}
\end{figure}

The gauge dependence of $Z_2$ in QCD can be formulated in a similar fashion.  
Analogous to (\ref{ex_l_qed}) the Lagrangian for QCD becomes
\begin{equation}
{\cal L}=-\frac{1}{4}F^2+\bar \psi\left(i\dsl{D}-m\right)\psi +\frac{\xi}{2} B^2 +B\partial 
\cdot A-\bar c\partial^\mu D_\mu c +\frac{\chi}{2}\bar c B
\label{ex_l_qcd}
\end{equation}
which is invariant under an extended BRS symmetry
\begin{eqnarray}
& &\delta^+ A_\mu=\epsilon D_\mu c\quad ,\quad \delta^+ \bar\psi=i\epsilon g c\bar 
\psi\quad ,\quad \delta^+ c=-\frac{1}{2} \epsilon g\left[ c,c\right]\nonumber\\
& &\delta^+ B=0\quad ,\quad \delta^+ \psi=-i\epsilon g c\psi \quad ,\quad \delta^+ \bar c=B
\label{qcd_x_brs}\\
& &\delta^+\xi=\epsilon\chi\quad ,\quad \delta^+\chi=0
\end{eqnarray}
The extended BRS symmetry (\ref{qcd_x_brs}) implies the following identity for the effective action
nearly identical in form to the QED identity (\ref{qed_gamma_x_brs}) 
\begin{equation}
0=\frac{\delta\Gamma}{\delta K_\mu}\frac{\delta \Gamma}{\delta A_\mu}
+ \frac{\delta\Gamma}{\delta \bar K} \frac{\delta \Gamma}{\delta \psi}
+  \frac{\delta\Gamma}{\delta K}
\frac{\delta \Gamma}{\delta \bar\psi}
+  B\frac{\delta \Gamma}{\delta \bar c}
+\frac{\delta\Gamma}{\delta \bar K_c}
+\chi\frac{\partial\Gamma}{\partial\xi}\frac{\delta \Gamma}{\delta  c}
\label{qcd_gamma_x_brs}
\end{equation}
where $K_\mu$ and $\bar K_c$ are currents coupled to composite operators respectively coupled to the
extended BRS variations of $A^\mu$ and $c$. Following the procedure used to 
develop (\ref{gamma_ident}) leads to a QCD expression in a similar form.
\begin{equation}
\frac{\partial}{\partial\xi}\frac{\delta^2\Gamma}{\delta\psi(y)\delta
\bar\psi(x)}  =
+
\frac{\delta^3\Gamma}{\delta\psi(y)\delta\bar K\delta \chi}
\frac{\delta^2\Gamma}{\delta\bar\psi(x)\delta\psi}
+
\frac{\delta^2\Gamma}{\delta\psi(y)\delta\bar\psi}
\frac{\delta^3\Gamma}{\delta\bar\psi(x)\delta K\delta\chi}
\label{qcd_gamma_ident}
\end{equation}
After transforming to momentum space we find a result identical in form to 
(\ref{prop_ident}).
\begin{equation}
\frac{\partial}{\partial\xi}S_F^{-1}(p)=S_F^{-1}(p)\left[ F(p,-p,0)+
\bar F(-p,p,0) \right]
\label{qcd_prop_ident}
\end{equation}
As in the QED case, we see that the necessary condition for gauge independence of $Z_2$ in QCD is for the 
Green function $F+\bar F$ to be zero on shell.  The distinction between 
QED and QCD occurs in the interactions, particularly the ghost-gluon interaction, which will contribute to $F(p,-p,0)$.
This is particularly evident at three loop level where diagrams (such as those in Figure 
\ref{non_ab_fig}) occur that cannot be related to the fundamental two- or 
three-point  Green functions.  Thus at three-loop level there is no simple extension of the result 
(\ref{on-shell_F_1}) 
from QED to QCD, and hence gauge independence of $Z_2$ in 
on-shell schemes seems problematic at the three-loop level and beyond in QCD.

\bigskip
\noindent
{\bf Acknowledgements:}  TGS is grateful for the
financial support of the Natural Sciences and Engineering Research Council of
Canada (NSERC).  TGS thanks Martin Lavelle and Emilio Bagan for discussions at early stages of this work.

\newpage

\end{document}